\documentclass[conference,compsoc]{IEEEtran}   	
\usepackage[pdftex]{graphicx}
\usepackage{textcomp}
\usepackage{authblk}

\usepackage{caption}
\usepackage{subcaption}
\usepackage{cite}
\usepackage{url}
\usepackage{siunitx}
\usepackage{booktabs}

\usepackage{algorithmic}
\title{Performance Evaluation of the Quorum Blockchain Platform}
\author{Arati Baliga, Subhod I, Pandurang Kamat and Siddhartha Chatterjee}
\affil{Persistent Systems Ltd.  \\
Bhageerath, 402 E., Senapati Bapat Road, Pune, India \\
Contact: blockchain@persistent.com}
\date{}							
\begin{document}
\maketitle
\IEEEpeerreviewmaketitle

\begin{abstract}
Quorum is a permissioned blockchain platform built from the Ethereum codebase with adaptations to make it a permissioned consortium platform.  It is one of the key contenders in the permissioned ledger space. Quorum supports confidentiality and privacy of smart contracts and transactions, and crash and Byzantine fault tolerant consensus algorithms.

In this paper, we characterize the performance features of Quorum. We study the throughput and latency characteristics of Quorum with different workloads and consensus algorithms that it supports. Through a suite of micro-benchmarks, we explore how certain transaction and smart contract parameters can affect transaction latencies. 
\end{abstract}

\section{Introduction}

Blockchain technology has progressed from the cryptocurrency realm to enterprise applications over the last two years and several enterprise blockchain and distributed ledger platforms have emerged to cater to this space.
Bitcoin \cite{bitcoin} first introduced and used blockchain as the underlying technology that powered the decentralized cryptocurrency. This version of blockchain is a permissionless, open network that provides cryptographically secured trust among untrusted entities to maintain a common distributed ledger. The ledger is secured by proof-of-work algorithm that is hard to subvert through collusion or otherwise.

Subsequently enterprise blockchain platforms emerged to cater to consortia of enterprises that needed permissioned, distributed ledgers for single-source of truth and data sharing. Permissioned blockchains can also be used in applications which currently require trusted intermediaries to perform mediating functions or attest to the integrity of data shared by multiple enterprises. These platforms are setup as a consortium of enterprises, have verified identities and admission control and secure the ledger using computationally more efficient (than proof-of-work) consensus algorithms. 

Blockchain platforms essentially encapsulate the low level implementation details about transaction and ledger organization, secure and robust consensus algorithms and provide a high level API that is easy to integrate for application developers. This enables speedy design and development of blockchain applications. In addition to the security and integrity of the distributed ledger, some of the key features enterprises care about are privacy of the transactions and data, transaction confirmation latency, overall throughput of the system and horizontal scaling of the blockchain data.

Some of the leading platforms  in the permissioned ledger space are Quorum \cite{quorum}, Hyperledger Fabric \cite{fabric}, CORDA\cite{corda}, Multichain \cite{multichain} and Chain \cite{chain}. While each platform has deployed their first production version ready to be used to build enterprise applications, there is almost no performance data available on any of them. To the best of our knowledge, this is the first paper that focuses on performance evaluation of Quorum.\\

In particular, this paper makes the following contributions: 

\begin{itemize}
\item We present an experimental evaluation of the Quorum 2.0 blockchain platform showing its performance characteristics.
\item Through a suite of micro-benchmarks custom built for Quorum, we study how different transaction and smart contract parameters can affect transaction latencies and their implications on application design.
\end{itemize}

This paper is organized as follows. In Section~\ref{sec:overview}, we provide an overview of the Quorum blockchain platform. In Section~\ref{sec:workloads}, we study the throughput and transaction latencies of the Quorum platform with controlled workloads. In Section~\ref{sec:microbenchmarking}, we describe the results of our microbenchmarking experiments and their implications for blockchain application design. We summarize related work in Section~\ref{sec:relatedwork} and finally conclude in Section~\ref{sec:conclusion}.

\section{Quorum Overview}
\label{sec:overview}
Quorum \cite{quorum} was developed by JP Morgan as a permissioned ledger implementation of Ethereum \cite{ethereum}.  Ethereum is a public permissionless blockchain that can be used across multiple domains to implement decentralized applications. It has support for Turing-complete smart contracts and therefore can be used to build general purpose blockchain applications across several domains. Being public and permissionless, its security is provided by the Proof-of-Work (PoW) consensus algorithm and its internal cryptocurrency Ether. PoW consensus algorithm adds deliberate cryptographic difficulty to prevent Sybil attacks on the Ethereum blockchain.  Its cryptocurrency Ether is consumed by all computations that run on the Ethereum blockchain thereby discouraging spam and denial of service attacks. Transactions in Ethereum are verified by all full-nodes in the network. Each node runs smart contracts within the Ethereum Virtual Machine (EVM) to verify transactions that invoke methods within the smart contract. 

Being a permissioned implementation of Ethereum, Quorum includes the following key changes to the Ethereum design:
\begin{enumerate}
\item \textit{Permissioned participation:} A permissioned membership implementation limits participation to a known set of nodes that have to be provisioned to be part of the blockchain network. Only these  are allowed to connect to the Quorum blockchain, verify transactions, run smart contracts and maintain the ledger state.
\item \textit{Consensus Algorithms:} By default Quorum provides RAFT \cite{raft-quorum} consensus for crash fault tolerance and IBFT \cite{ibft} consensus for Byzantine fault tolerance. These provide a replacement for Ethereum's PoW implementation. While PoW protects a public blockchain by deliberately introducing cryptographic difficulty, it is unnecessary and wasteful (excessive power consumption) in a permissioned setting where participants are known. RAFT and IBFT algorithms lead to faster consensus and provide immediate transaction finality, making them a suitable choice for permissioned blockchain implementations. Quorum also supports a pluggable architecture where a different consensus implementation can be plugged in if needed. 
\item \textit{Support for privacy:} Privacy was one of the design goals in Quorum. It allows subsets of parties in a larger consortia to transact with one another without making the transactions public to members of the larger consortia. Privacy is enabled in Quorum by splitting the larger public ledger into a public and a private ledger. The public ledger (and the associated state) is visible to all the nodes in the network, while the private ledger (and associated state) is visible only to the transacting parties. Only a hash of the private transaction appears on the public ledger and is visible to other nodes that are not counterparties to the transactions. Only the counterparties to the private transaction have the keys to decode and view it. Similarly smart contracts can also be deployed privately and will be visible only to the transacting parties. A detailed explanation of private transaction process flow can be found online \cite{privatetxflow}.
\item \textit{Elimination of transaction pricing:}  Quorum eliminated the concept of adding cost to a transaction using gas. Therefore, Quorum does not have any cryptocurrency costs associated with running transactions on the Quorum network.  Since the Quorum code was intially forked off Ethereum,  the usage of gas itself exists but is set to zero in the Quorum implementation. Gas is used by Ethereum to pay for transactions (in ethers) to incentivize miners and protect the Ethereum network from spam and DDoS attacks. 
\end{enumerate}

\begin{figure}[!thbp]
    \centering
    \includegraphics[width=0.9\linewidth]{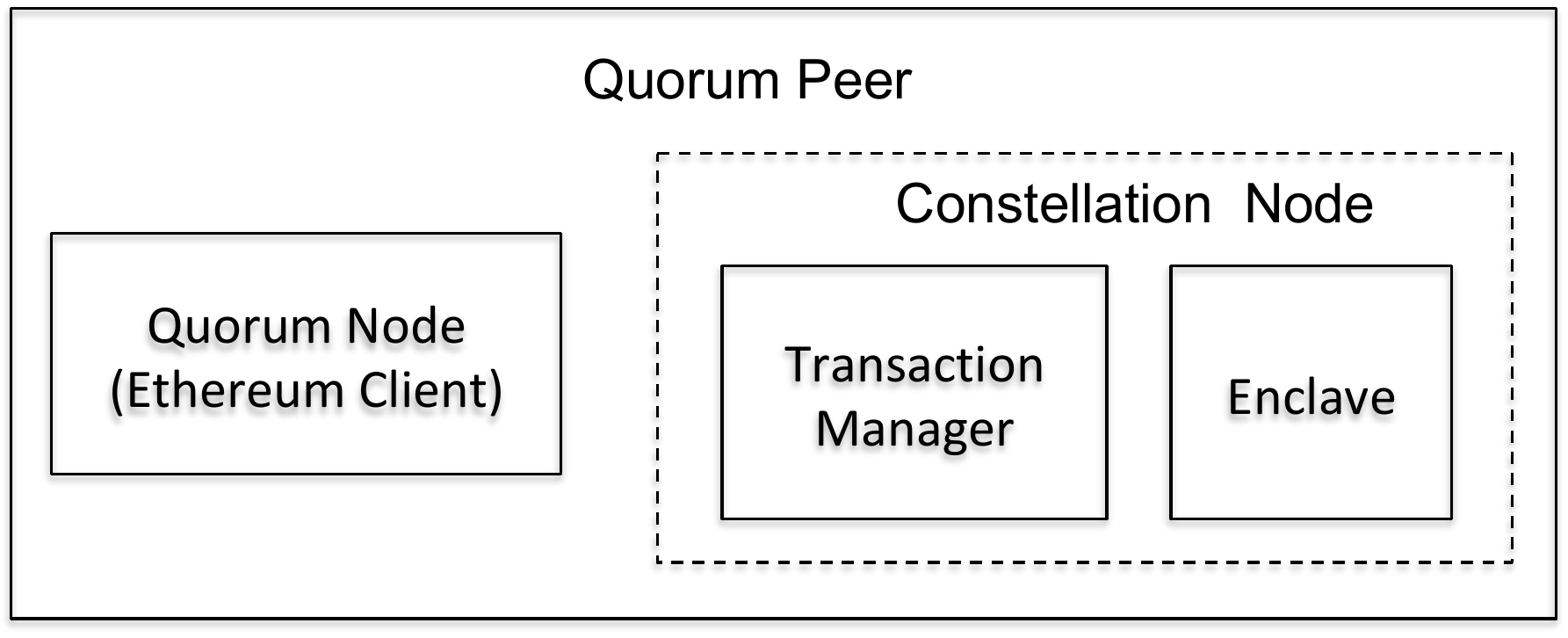}
    \caption{Quorum Architecture}
    \label{fig:quorum-architecture}
\end{figure}

Figure ~\ref{fig:quorum-architecture} shows the architecture diagram of a Quorum node. The code itself is forked from \textit{geth} (the Ethereum Go client) and modified to work in a permissioned setting as discussed previously.  It manages the communication with clients as well as other nodes using the general-purpose Constellation \cite{constellation} p2p system for communicating with secure messages.  The Constellation module on the Quorum node comprises of two sub-modules - Transaction Manager and the Enclave. The Transaction Manager manages private transactions by enabling access to them and sending and receiving encrypted payloads to other Transaction Managers running on other Quorum nodes. It leverages the Enclave for performing cryptographic operations. The Enclave acts as an independent module that seals all the private keys belonging to the transactions. All encryption/decryption operations are done within the Enclave.

Quorum supports the RAFT and IBFT Consensus algorithms.  Older version of Quorum supported QuorumChain, a basic m-of-n voting based consensus. It is no longer supported in the current version.\\

\textbf{RAFT:} Raft is a consensus algorithm used for managing replicated state machines or logs. It is equivalent to the PAXOS \cite{paxos} algorithm proposed for building crash tolerant systems in terms of its fault tolerance properties and performance but is simpler to understand and implement. Quorum uses etcd's Raft implementation \cite{raft-quorum}. This is useful for consortia where there the consotia members are known and provisioned into the system. A Leader is responsible for generating new blocks. RAFT offers faster block times and does not create unnecessary empty blocks. RAFT need 2f+1 nodes to be setup in the network to have the capability to tolerate f faulty nodes. A complete description of the RAFT consensus algorithm can be found in the RAFT paper \cite{raft}.

\textbf{Istanbul BFT(IBFT):} Istanbul BFT is a Byzantine fault tolerant state machine replication based consensus algorithm. It is modeled after Castro and Liskov's Practical Byzantine Fault Tolerance (PBFT) algorithm \cite{pbft} . It is also based on 3-phase commit like PBFT and uses the PRE-PREPARE, PREPARE, and COMMIT stages. Before each round, the nodes will pick one of them as the Leader(Proposer). Proposer is responsible for proposing new blocks in the network. In each state I.e. PRE-PREPARE, PREPARE and COMMIT, validators broadcast the State message and wait for 2f+1 State messages. Upon receiving 2f+1 state messages Validators commit the current state and move forward to next state. The system can tolerate at most of f faulty nodes in a network with  3f + 1 nodes.

\section{Characterizing Latency and Throughput}
\label{sec:workloads}
In this section, we summarize our findings on transaction throughput and transaction latency measurements that we conducted on the Quorum network using RAFT and IBFT consensus algorithms. 

\subsection{Metrics}
\textit{Transaction throughput} is defined as the number of transactions per second successfully processed by the blockchain network.  A transaction is successfully processed when it is included in a block and committed as part of the ledger. \textit{Transaction latency} is the time elapsed between when a request is sent, to the time when the response is received by the client. For read transactions, it is the time taken to receive the response for a read query. For write transactions, it is the time elapsed between the request and an event confirmation as received by the client after the transaction is confirmed on the blockchain.  

\subsection{Experimental Setup}
\label{sec:experimentalsetup}
A private blockchain network was setup with three peers when using the RAFT consensus algorithm and with four peers when using the IBFT consensus algorithm (the minimum configuration requirement for both respectively). All peers were run on hardware machines within our network. Each machine had 8 vCPUs (4 cores at 3.6 GHz with hyperthreading) and 16 GB RAM. We used three more machines with the same configuration to run the clients. All nodes had the Ubuntu 14.04 LTS operating system and were connected to each other with a 1 Gbps switch.
\label{sec:hardwaresetup}

\begin{figure*}[!tphb]
\centering
\begin{subfigure}{.57\linewidth}
  \centering
  \includegraphics[width=\linewidth]{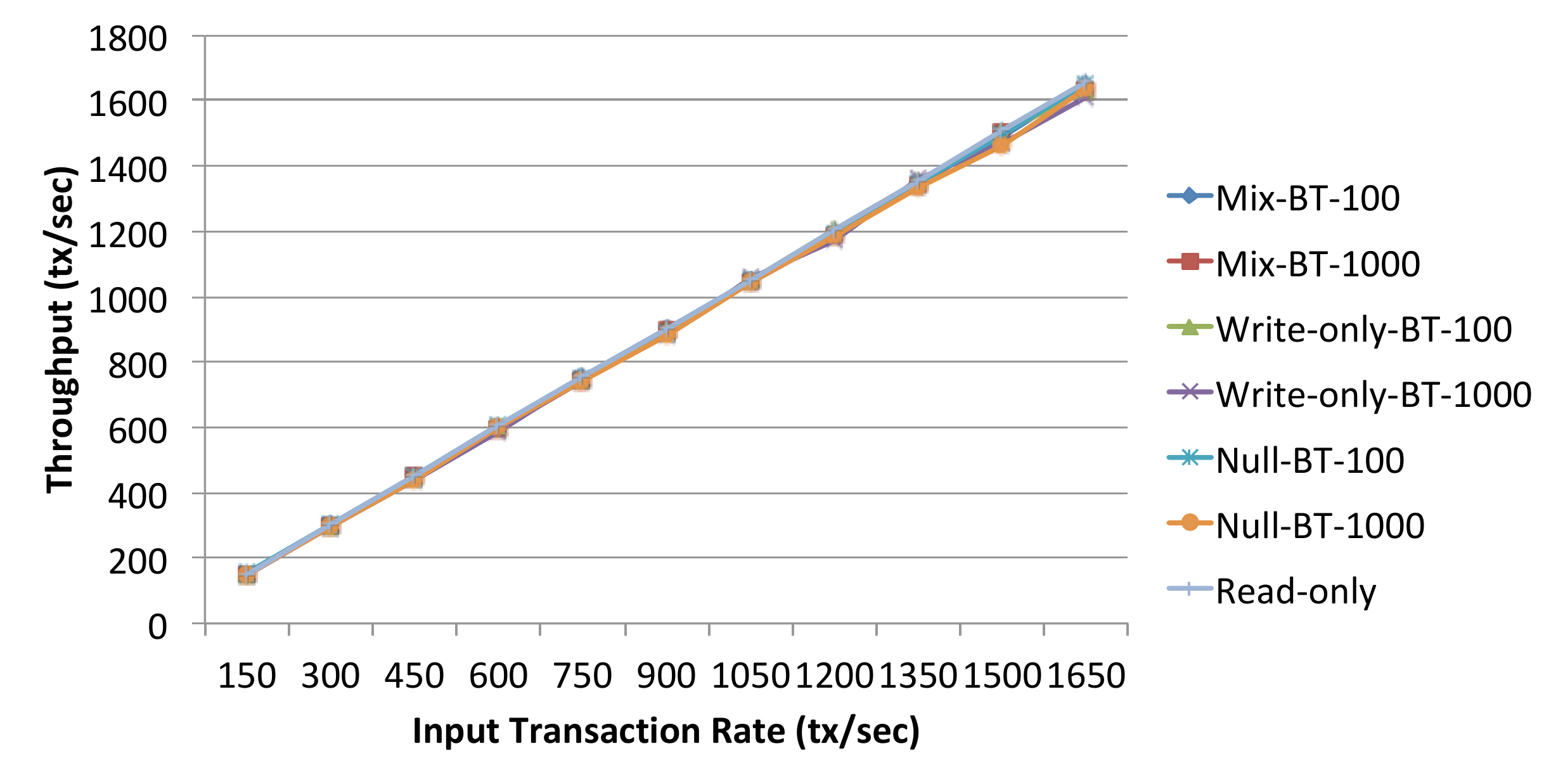}
  \caption{Transaction throughput}
  \label{fig:raftthroughput}
 \end{subfigure}%
\begin{subfigure}{.43\linewidth}
  \centering
  \includegraphics[width=\linewidth]{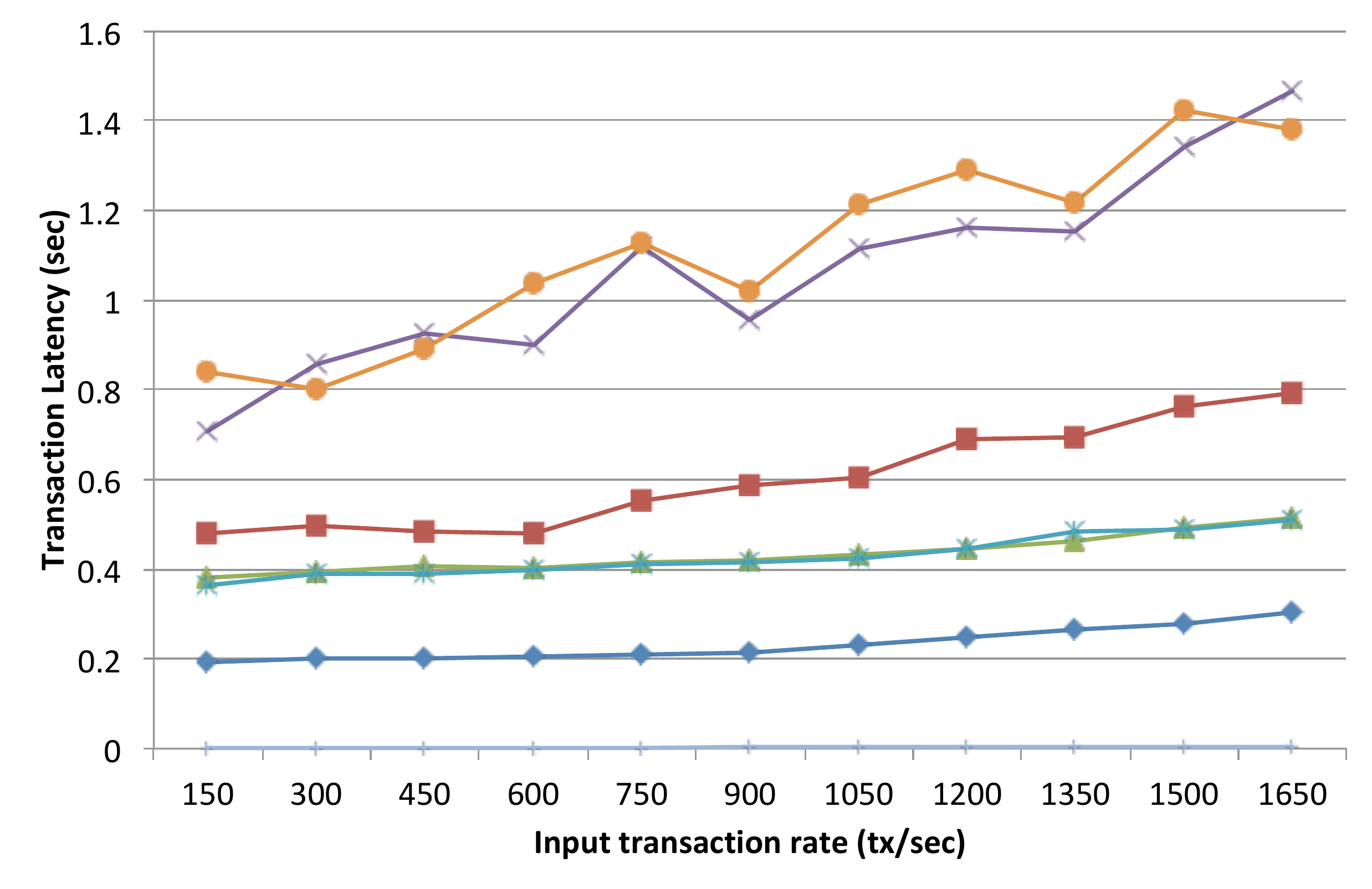}
  \caption{Transaction latency}
   \label{fig:raftlatency}
\end{subfigure}
\caption{Latency and throughput measurements with RAFT consensus}
\label{fig:raft}
\end{figure*}

\subsection{Client Setup}
\label{sec:clientsetup}
We extended the Caliper benchmarking tool \cite{caliper} developed by Huawei Technologies \cite{huawei}  by adding a Quorum plugin to it. Caliper is now officially incubated as a Hyperledger project \cite{hyperledger-caliper}. Caliper by default works for Hyperledger Fabric \cite{fabric} and Hyperledger Sawtooth Lake \cite{sawtoothlake}. The Quorum plugin enables Caliper to send controlled workloads to the Quorum network to record its throughput and transaction latencies using Caliper's measurement framework. Caliper runs on the client machines and sends transactions to peers in the Quorum network. Each client in our setup sends controlled workloads to a different peer to balance the workload across the network. Transactions are considered confirmed on the blockchain when the Quorum peer emits a block event indicating inclusion of the transaction in the block. Caliper listens to block events and calculates resulting throughput and latencies on the client.

Apart from adding the Quorum plugin to enable Caliper to interact with Quorum, we made another major change to the Caliper framework. Caliper client was missing out on certain block events at higher transaction rates resulting in failed transactions. This was because the Caliper client itself was single threaded and was performing both functions of sending transactions as well as listening to events resulting from transaction confirmations. We modified the client to spawn a new process that essentially splits the listener and processor into separate processes. A newly spawned process only listens to block events and inserts them in a messaging queue to be processed later by Caliper's main process. By having a separate process listen continuously to block events, we have completely eliminated occurrence of failed transactions due to missed block events at higher transaction rates. 

\subsection{Load Generation}
Caliper is run on all the client machines. Each client sends transactions to a different peer on the Quorum network. 
In each experiment the send rate is varied starting from 50 transactions per second to 550 tx/sec, which was the maximum capacity for client nodes used in our experiments. The total load on the Quorum network ranges from 150 tx/sec to 1650 tx/sec. Each client sends transactions at a specified send rate, halts for 5 secs,  and starts the
next round. The experiment is repeated for three rounds.  At the end of the third round, an average of the throughput and latencies is calculated. The blockchain network is subject to a maximum of 49500 total transactions. We also monitor the CPU utilization and memory consumed on the peers for the duration of the experiment. 

\subsection{Workloads}
For all workloads, a smart contract is deployed and pre-loaded with key-value pairs. We used the following workloads.
\begin{itemize}
\item \textit{Write-only workload:} The write-only workload comprises of all write transactions that update a value for a randomly selected key in the key-value store of the smart contract. Write generates a transaction on the blockchain that requires the consensus algorithm to execute successfully.
\item \textit{Null workload:}   The null workload comprises of transactions that call a function within the smart contract that simply returns. The null function skips the processing within the smart contract and therefore represents the baseline cost for the write call. 
\item \textit{Read workload:} The read workload comprises of read transactions that read the values for randomly selected keys from the key-value store within the smart contract. Read workload is generated by all clients sending their transactions to a single peer. This design is intentional as reads are served locally by the peer by performing lookups within its local data store. Reads do not generate a transaction on the blockchain.
\item \textit{Mix workload:} Mix workload has a 50-50 mix of reads and writes. 
\end{itemize}

\subsection{Throughput and Latency Measurements}
To evaluate the latency and throughput of the system, it is important to understand how the BlockTime parameter affects the overall latency and throughput of the system.  BlockTime parameter provisioned during network setup controls when transactions are batched into blocks.

\begin{table}[htpb]
\centering
\begin{tabular}{ |c|c|c|c|c|c| } 
 \hline
{ } & \multicolumn{5}{c|}{Block Time (ms)} \\
{}  & 50 & 100 & 250 & 500 & 1000 \\ 
  \hline
Throughput (tx/sec) & 752  &  752 & 750 & 747 & 748 \\ \hline
Latency  (secs) & 0.463 &  0.414 & 0.533 & 0.589 & 1.006\\ \hline
\end{tabular}
\caption{Effect of RAFT block time on throughput and latency with input transaction rate of 750 tx/sec.}
\label{tab:blocktime}
\end{table}

\subsubsection{Tuning Block time}
In Quorum, with RAFT consensus, the default block time is set to 50 ms. To understand how block time affects throughput and latency, we recorded latency and throughput with different block time settings as shown in Table~\ref{tab:blocktime} with a moderate input transaction rate of 750 tx/sec.  It is seen from the results that the throughput is more or less constant and is not affected by the block time setting \footnote{Note that the resulting throughput numbers are slightly higher than the input transaction rates in some cases. This is because of transaction confirmations clustering together, shortening the time window between the first and last transaction confirmed leading to a slightly higher throughput number.}. The transaction latencies however increase when the block time is increased. The latencies increase due to the increased time needed to include the transaction in the block. Therefore block time can be tuned based on latency sensitivity of applications. For IBFT, the default block time setting is 1 sec, which is the minimum block time setting available.  
\begin{figure*}[!tphb]
\centering
\begin{subfigure}{.47\linewidth}
  \centering
  \includegraphics[width=\linewidth]{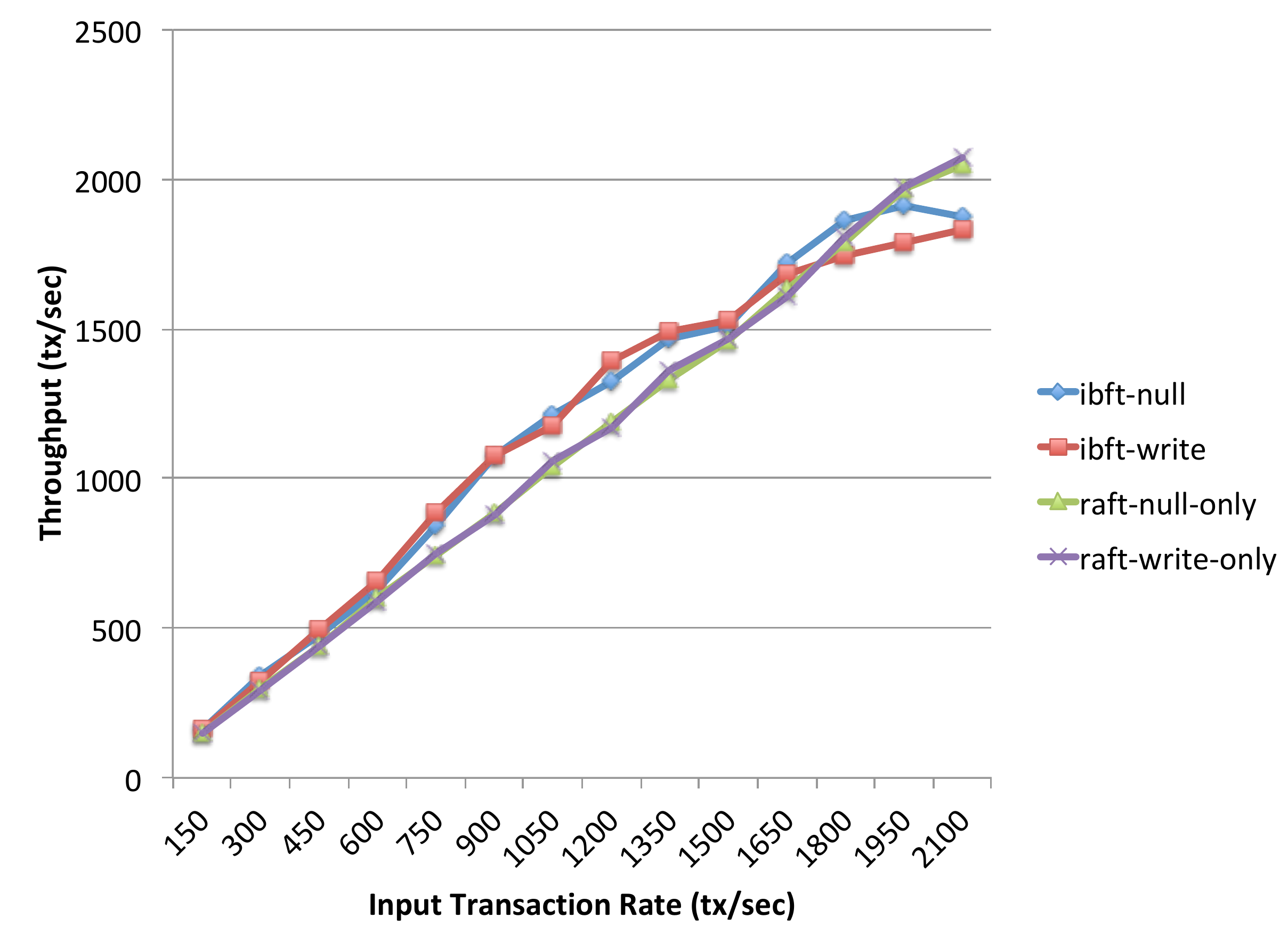}
  \caption{Transaction throughput}
  \label{fig:raftibftthroughput}
 \end{subfigure}%
\begin{subfigure}{.47\linewidth}
  \centering
  \includegraphics[width=\linewidth]{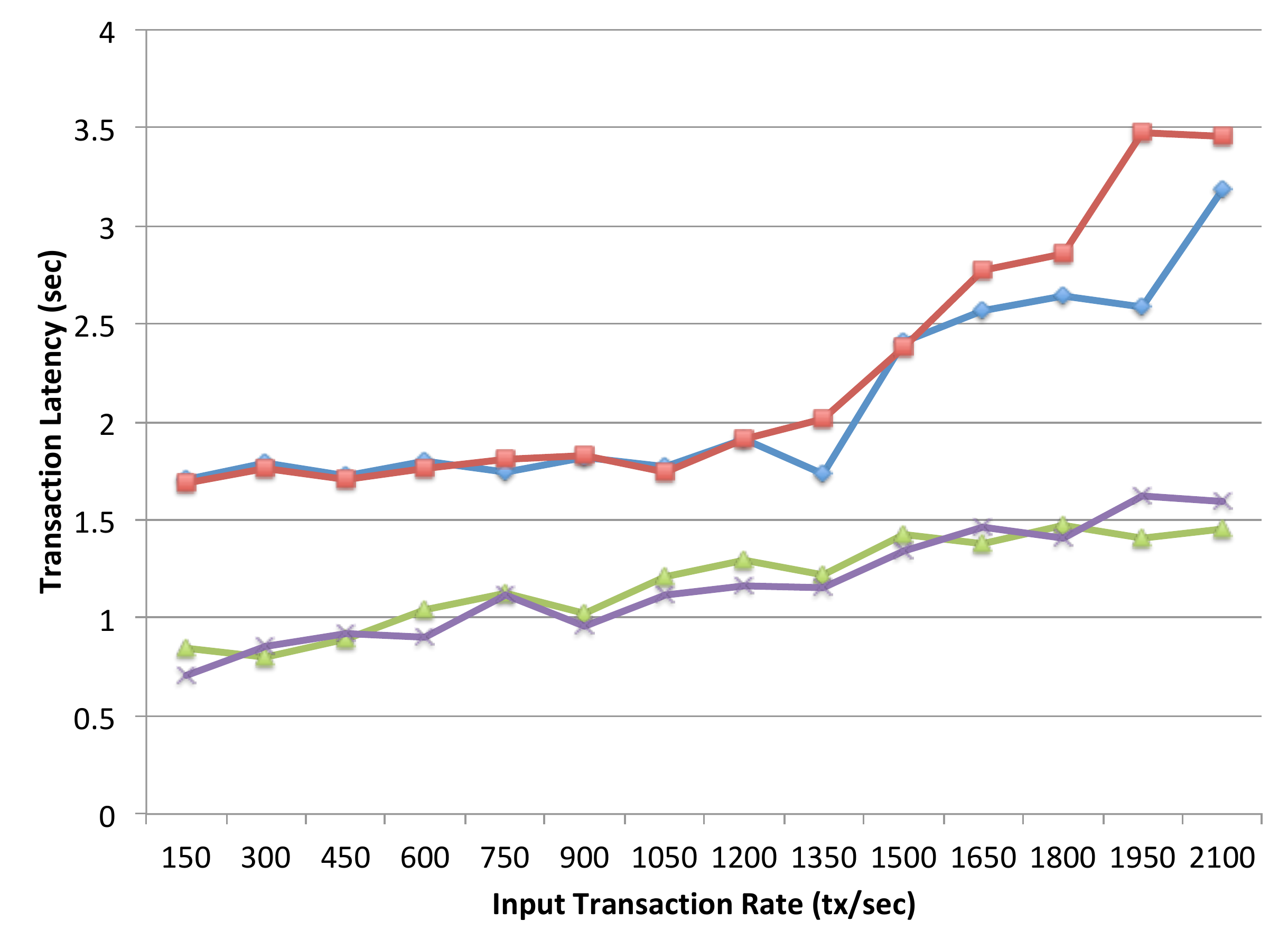}
  \caption{Transaction latency}
   \label{fig:raftibftlatency}
\end{subfigure}
\caption{RAFT versus IBFT consensus - Throughput and Latency measurements}
\label{fig:raftibft}
\end{figure*}

 \begin{figure*}[!tphb]
\centering
\begin{subfigure}{.5\linewidth}
  \centering
  \includegraphics[width=\linewidth]{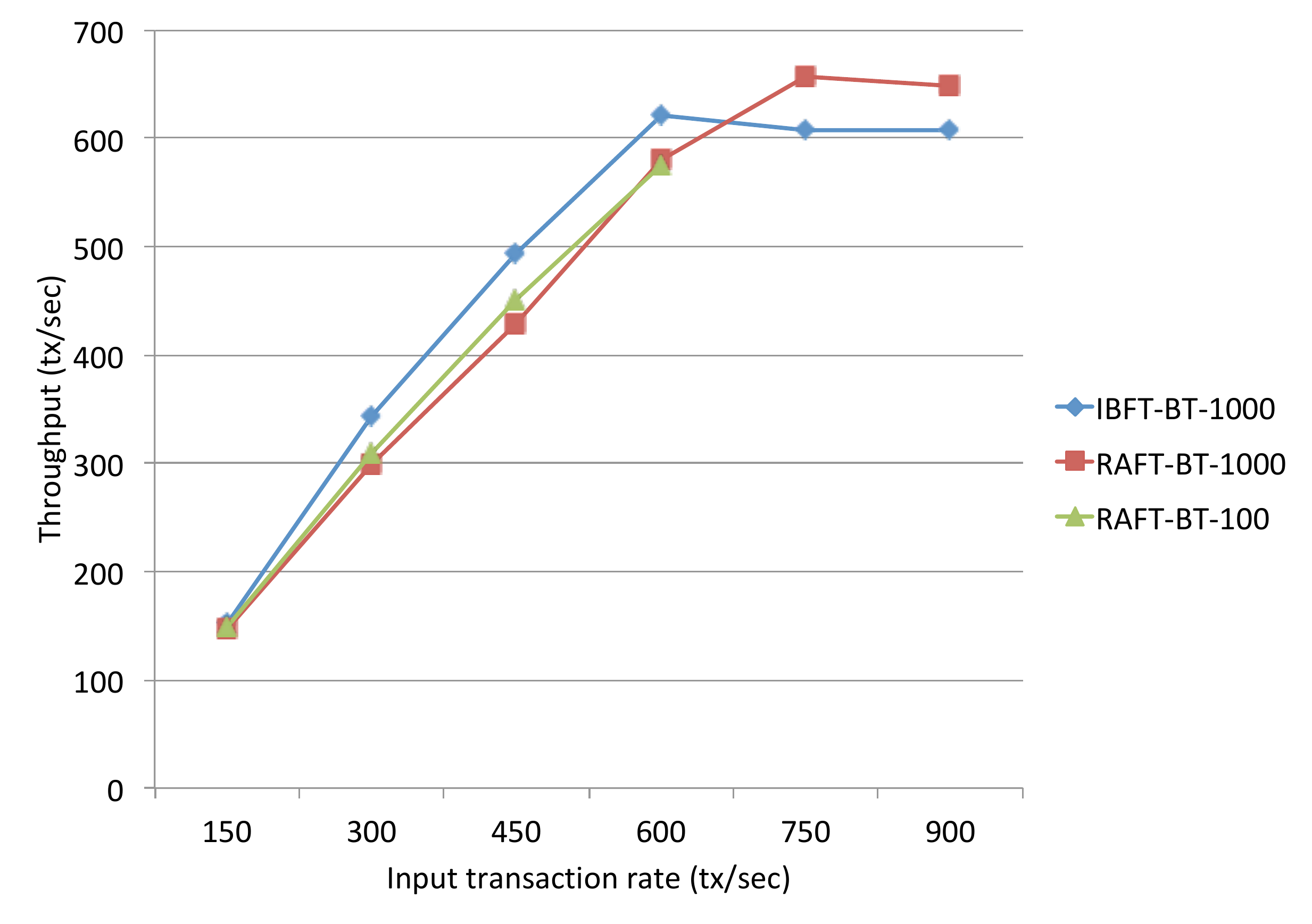}
  \caption{Transaction throughput}
  \label{fig:pvtcontractthroughput}
 \end{subfigure}%
\begin{subfigure}{.5\linewidth}
  \centering
  \includegraphics[width=\linewidth]{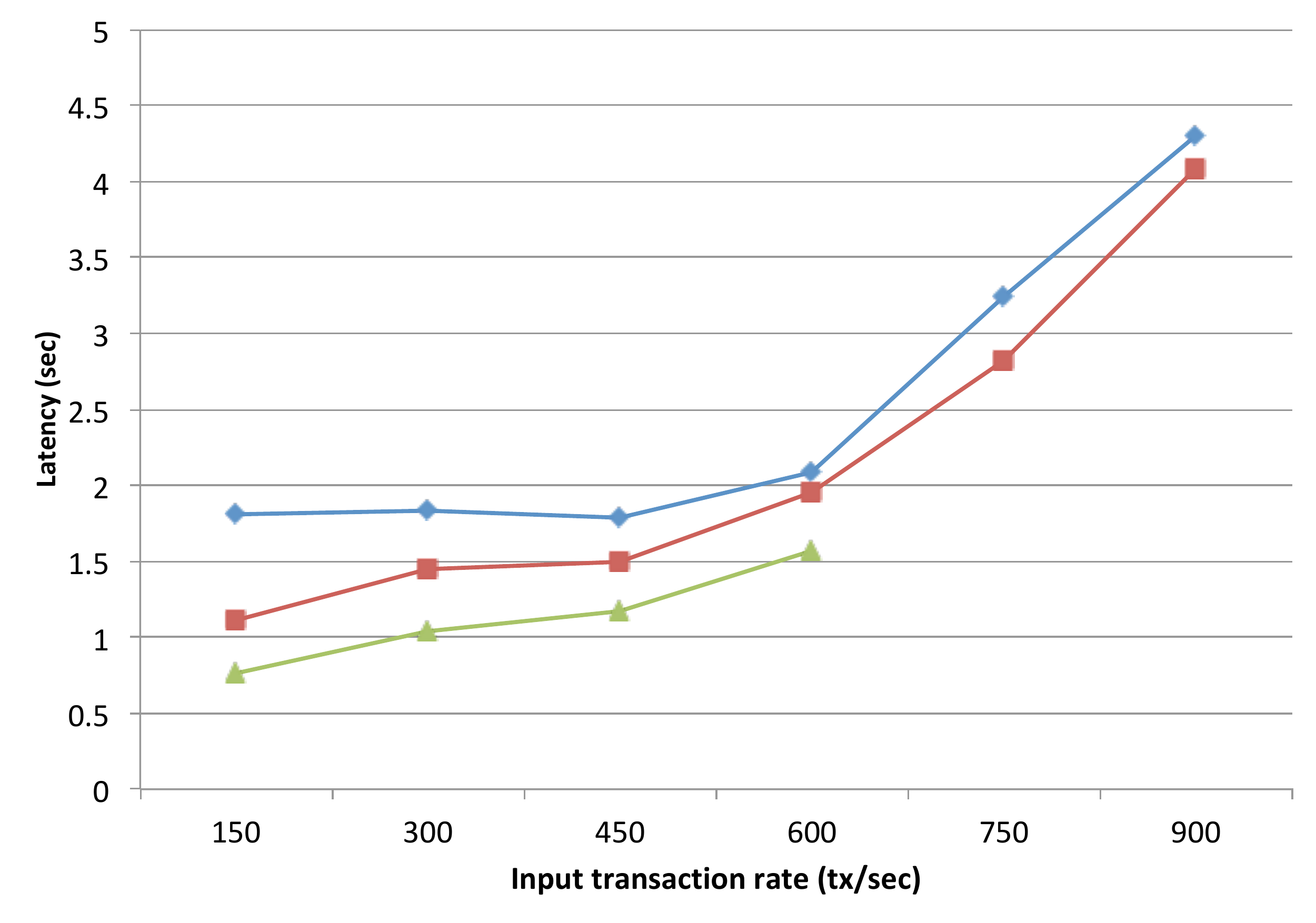}
  \caption{Transaction latency}
   \label{fig:pvtcontractlatency}
\end{subfigure}
\caption{Latency and throughput measurements for  private contract deployments}
\label{fig:pvtcontract}
\end{figure*}

\subsubsection{Latency and Throughput with RAFT Consensus}
RAFT is the default consensus algorithm used with Quorum for systems that do not need Byzantine fault tolerance. 
Figure~\ref{fig:raft} shows the latency and throughput measurements for all the four workloads. Each workload except for the read-only workload, has two plots. One with BlockTime 100ms (shown as BT-100) and the other with BlockTime 1000ms (shown as BT-1000) in the figure.  The maximum load of 1650 tx/sec was generated on the blockchain network with three clients, each generating a load of upto 550 tx/sec.

The results show that for the given transaction rates, all workloads scale linearly for the entire range indicating that Quorum has good scaling characteristics.  The latencies however show interesting variations for different workloads. As expected, the read workload exhibits the least latency as it is a local lookup from the Quorum peer's key-value store and does not involve transaction ordering and consensus.  Null and write workload latencies are mainly dependent on the block time parameter setting.  Both curves follow each other very closely indicating that the actual time taken for performing the update operation is very small. It is important to note here that the write workload has transactions that update only a single entry in the smart contract key-value store. The write and null workloads with the block time of 1000 ms have the highest latencies. The mix workload has a 50-50 mix of reads and writes and therefore falls appropriately in the expected latency range.

\subsubsection{RAFT versus IBFT}
IBFT can be used by Quorum deployments that need Byzantine fault tolerance. These set of experiments are designed to compare the performance of the IBFT deployment with RAFT deployment. To make them comparable, we use a block time of 1 sec for both RAFT and IBFT.  Block time of 1 sec is the minimum possible setting for the IBFT algorithm. The experimental setup uses one extra client for RAFT to generate loads on the network above 1650 tx/sec. Figure~\ref{fig:raftibft} shows the throughput and latency comparison between RAFT and IBFT.  

It is seen from the results that IBFT actually provides slightly higher throughput up until an input load of 1500 tx/sec, which is contrary to the expectation. RAFT starts to perform slightly better when the load is increased beyond 1650 tx/sec. Both algorithms scale quite well  with RAFT gaining a slight advantage for very high input transaction rates. The transaction latencies however are significantly higher for IBFT consensus. For  most data points IBFT latencies are almost double or more than double despite both RAFT and IBFT having the same block time setting of 1 sec.

\subsubsection{Public versus Private Contracts}
Private contracts in Quorum involve additional encryption/decryption operations and secure communication overhead between peers. This feature was built to support transaction privacy where a subset of parties transacting with each other within a large consortium can do so without others having any knowledge of the private transactions. In this case, the privately transacting parties will have to deploy a private smart contract that will be stored and run only on the peers that are counter-parties to the transaction. The goal of this experiment was to measure the throughput and latency of the system when private contracts are deployed instead of public contracts. Our controlled workloads described so far use public smart contracts. 

For these set of experiments, we deployed a private contract between all the peers. Latency and throughput is measured by using the write workload which invokes methods within the private contract.
Figure~\ref{fig:pvtcontract} shows the throughput and latency measurements for the write workload with RAFT and IBFT consensus. RAFT experiments were conducted with Block time of 100 ms and 1000 ms. 

The results show that the throughput is comparable for all settings with public contracts up until an input transaction rate of 600 tx/sec. If the load is increased further, the throughput starts to degrade. This is probably due to the increased messaging overhead between the nodes and the cryptographic operations involved in encrypting/decrypting confidential transactions. 
We also found that for the RAFT experiment with the block time of 100 ms, when the input transaction rate is increased beyond 600 tx/sec, errors are encountered. Upon further investigation, we realized that it is a bug in the Quorum code pertaining to inappropriate handling of file descriptors, where the code eventually runs into not enough file descriptors available scenario. Assigned file descriptors are allocated but not correctly released in the current version of the Quorum code.  While we have reported this issue to the Quorum team, the fix is pending and therefore, we were unable to conduct further experiments with higher loads using this setting. The maximum load that private contracts could handle was 900 tx/sec. Increasing the load beyond this point leads to consensus failure.

\section{Micro-benchmarking Experiments}
\label{sec:microbenchmarking}
In these set of experiments, we use three custom-built micro-benchmarks that study other parameters that affect transaction latencies. The micro-benchmarks measure latencies by altering the number of reads and writes performed by transactions, conducting reads and writes on different size key-value stores and transacting with different sized transaction and event payloads. 
For each of the experiment described below, we run the micro-benchmarks and record transaction latencies averaged across a large number of runs. These experiments used only a single client to submit transactions and a single peer. The smart contract generates an event to notify the client of completion of the operation. End-to-end latencies are measured by sending one transaction at a time. All micro-benchmarks use the RAFT consensus with a block time of 1 ms\footnote{Block time of 1 ms is the minimum setting for RAFT. We use the minimum setting to prevent the block time from factoring into and/or (dominating) the actual transaction latency.}.

\subsection {Read Set and Write Set Size}
\label{sec:rwsetsize}

This micro-benchmark measures transaction latencies by varying the number of key-values read (read-set) and the number of key-values written (write-set). The goal is to understand how the size of the read-set and the write-set affects transaction latencies. The key-value store within the smart contract is initialized prior to the experiment. 
The client invokes a  smart contract method causing it to generate the desired numbers of reads or writes.

\begin{figure}[!thpb]
    \centering
    \begin{subfigure}{.9\linewidth}
    \centering
    \includegraphics[width=\linewidth]{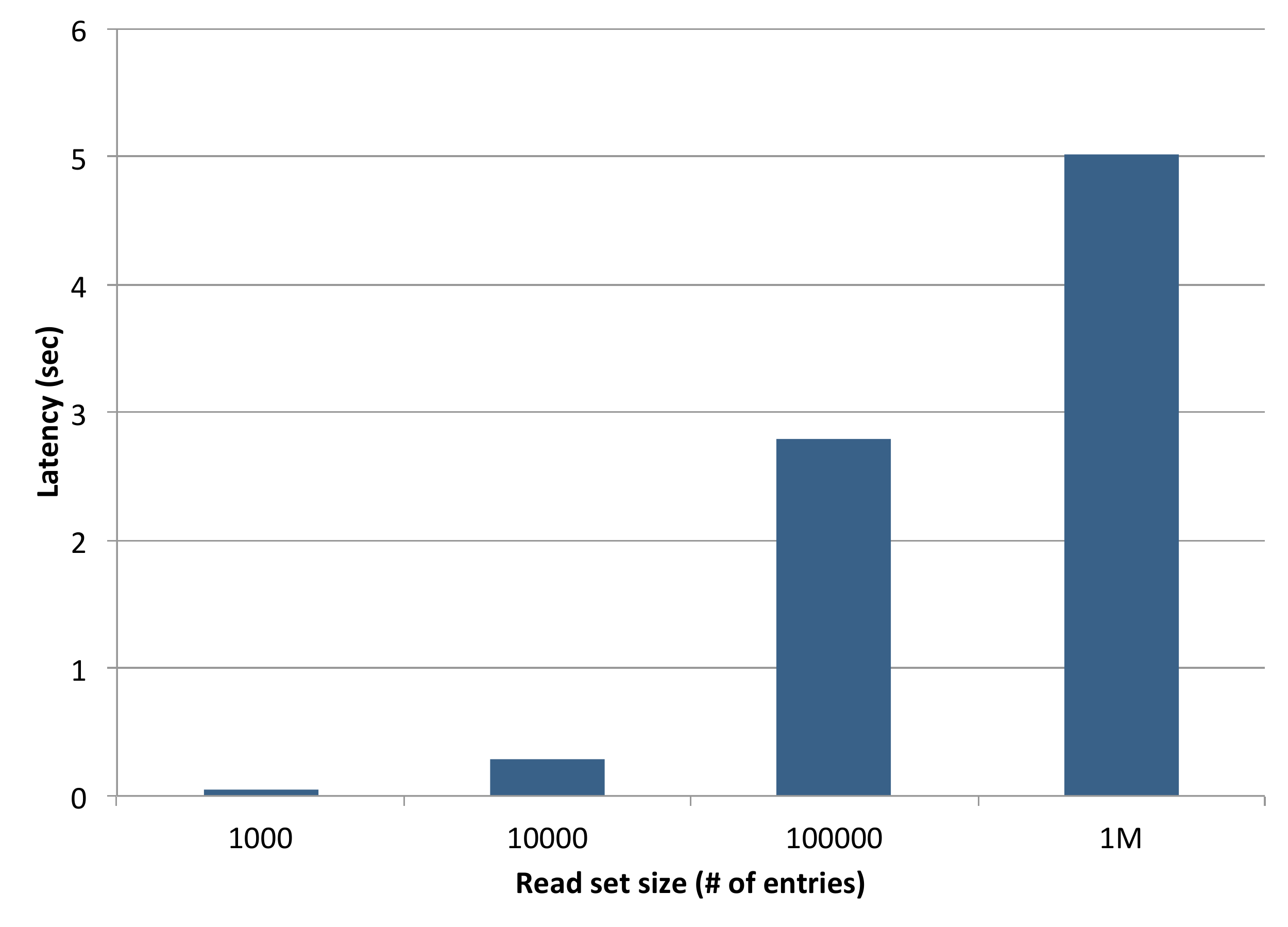}
    \caption{Read latencies}
    \label{fig:rwset-readlatency}
    \end{subfigure}
    \begin{subfigure}{.9\linewidth}
    \centering
     \includegraphics[width=\linewidth]{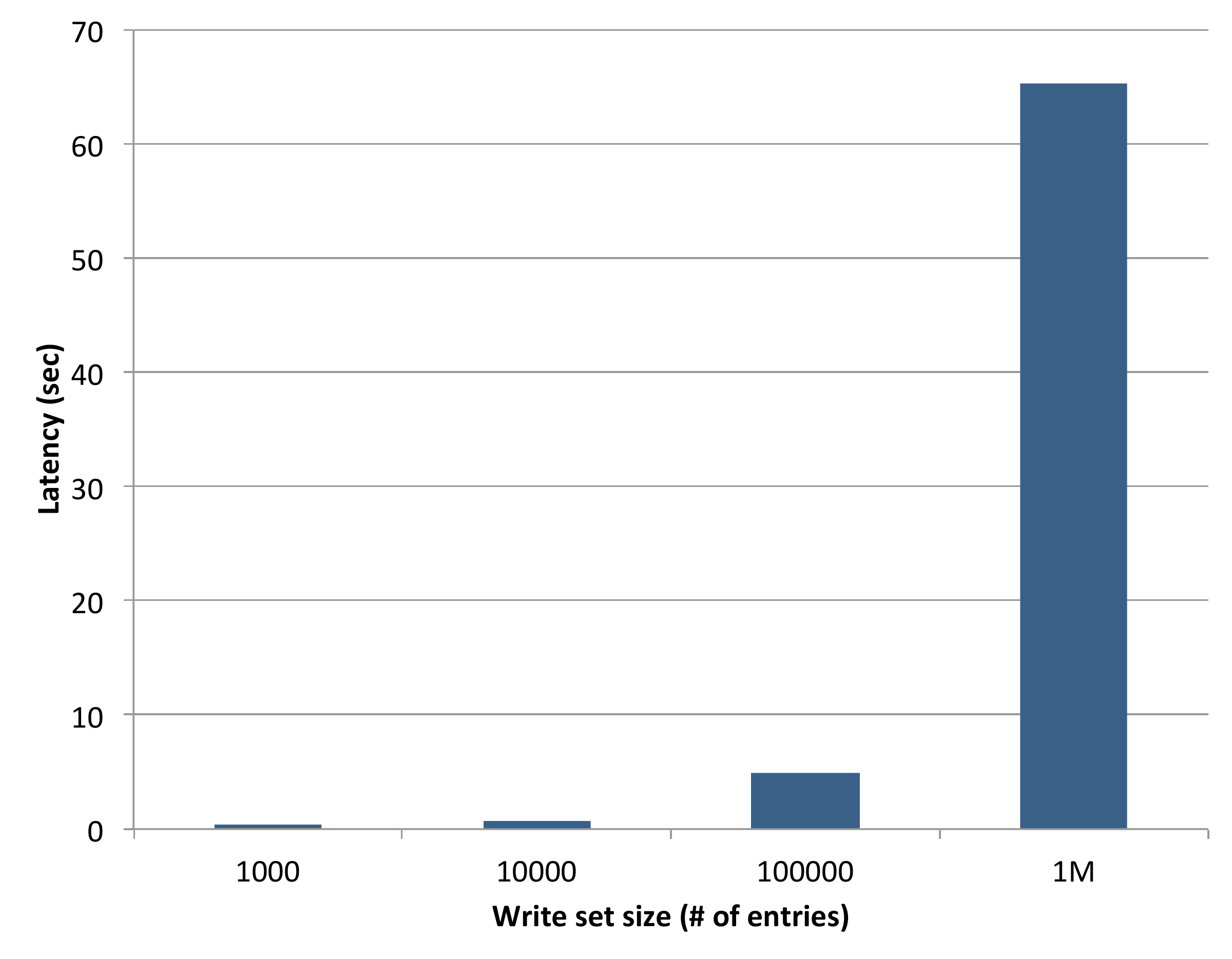}
    \caption{Write latencies}
    \label{fig:rwset-writelatency}
    \end{subfigure}
    \caption{Read and write latencies for varying read-set and write-set sizes}
\label{fig:rwsetsize}
\end{figure}

Figure~\ref{fig:rwsetsize} shows the latencies increase with increasing size of read-set and write-set as expected.  The total latency however increases significantly when 1 million entries are written while the increase is not as sharp when 1 million entries are read. We were unable to carry out the experiment by further increasing the size of the write-set as it led to a failure in the RAFT consensus mechanism \cite{raft-failure}. This is a current limitation in Quorum.

\begin{figure}[!thpb]
    \centering
    \begin{subfigure}{.9\linewidth}
    \centering
    \includegraphics[width=\linewidth]{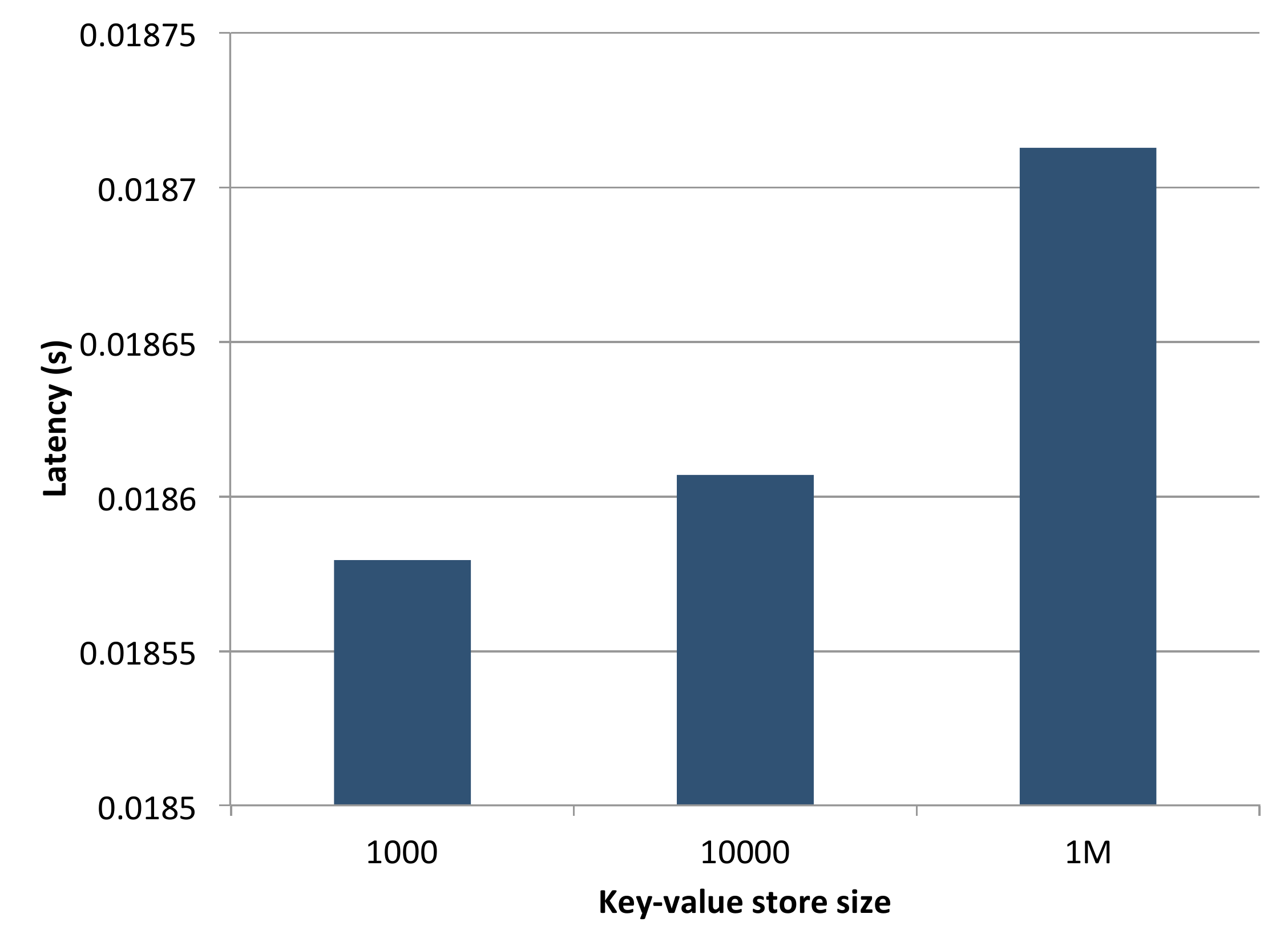}
    \caption{Read latencies}
    \label{fig:kvstore-readlatency}
    \end{subfigure}
    \begin{subfigure}{.9\linewidth}
    \centering
     \includegraphics[width=\linewidth]{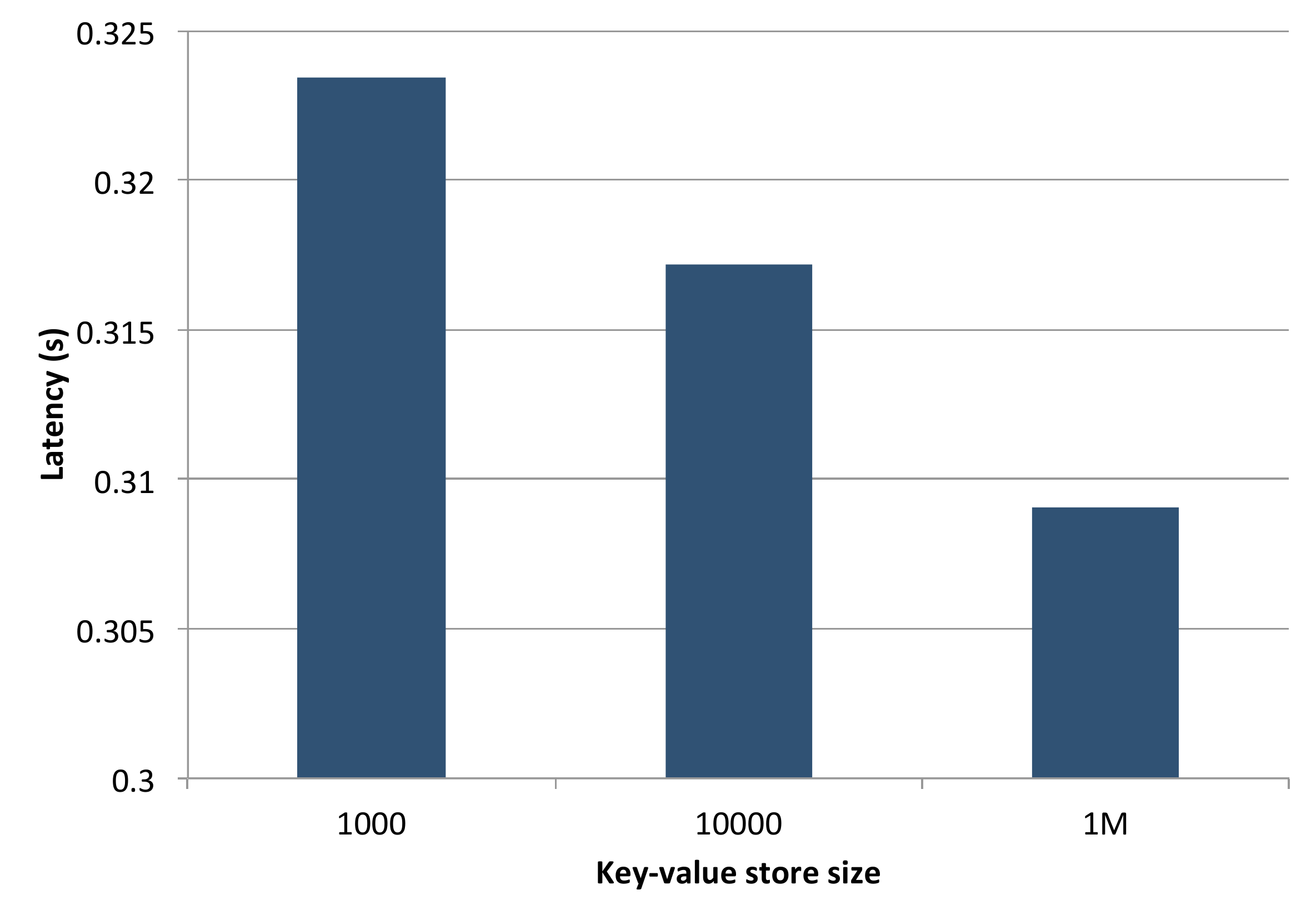}
    \caption{Write latencies}
    \label{fig:kvstore-writelatency}
    \end{subfigure}
    \caption{Read and write latencies for varying sizes of key-value store}
\label{fig:kvstore-rwlatency}
\end{figure}

\subsection {Smart contract Key Value Store Size}
\label{sec:keyvaluestoresize}
This micro-benchmark measures the transaction latencies when reads and writes are performed with different sizes of key-value stores. In real applications that store application related data, the smart contract key-value store might grow over a period of time. This would essentially tell us if the read-write latency would degrade when the size of the key-value store increase significantly over time.
We pre-populated smart contract data ranging from 1000 entries to 1 million entries in different experiments. The reads and writes were uniformly spread across the entire data set. The experiment was repeated for different key-value store sizes.  

Figure ~\ref{fig:kvstore-rwlatency} shows that reads and writes are highly optimized and relatively unaffected by the key-value store size up to 1 Million entries.  

\subsection{Transaction and Event Payload Sizes}
\label{sec:payloadsizes}
This micro-benchmark measures the latencies by varying the transaction and event payload sizes. Transaction payload is the payload passed to the smart contract method when invoking it. In Quorum, the maximum transaction payload size is 32KB. Therefore, we use payload sizes in increments of 10 KB to see their effect on transaction latencies. Event payload is the payload passed from the smart contract to the event listener.  This is a separate experiment which sends different sizes of event payloads. 

\begin{table}[hbpt]
\vspace{0.25cm}
\centering
\begin{tabular}{ |c|S[table-format=3.3]|S[table-format=3.3]|} 
 \hline
 {} &  \multicolumn{2}{c|}{Latencies (s)}\\ \hline
{Payload Size(KB)}  & {Tx Payload} & {Event Payload} \\ 
  \hline
  \hline
1 & 0.325  & 0.330 \\ \hline
10 & 0.383  &  0.373 \\ \hline
20 & 0.384 &  0.395 \\ \hline
30 & 0.407 & 0.404  \\  \hline
\end{tabular}
\caption{Transaction latency with variable sized transaction and event payloads}
\label{tab:payload}
\end{table}

Table~\ref{tab:payload} summarizes the findings in both sets of experiments.
The results show that the transaction latencies increase with each 10KB increase in the payload.  The transaction latency increases by 25.23\% when the transaction payload size is increased from 1 KB to 30KB and the transaction latency increases by 22.4\% when the event payload is increased from 1KB to 30KB. Therefore, application designers need to take that into consideration when designing transaction and event payloads.

\section{Related Work}
\label{sec:relatedwork}
Blockbench  is a framework proposed for benchmarking the performance of private blockchain platforms \cite{blockbench}. In the Blockbench paper, the focus is  on performance comparison of Fabric (v 0.6), Ethereum and Parity \cite{parity}. Some performance results are available recently from a paper published by the developers of Hyperledger Fabric \cite{fabricpaper}. To the best of our knowledge, this is the first paper that does a thorough  performance evaluation of the Quorum blockchain platform. 

The Caliper tool \cite{caliper} was developed by Huawei Technologies to measure the throughput and latency of permissioned blockchain platforms. We have built a Quorum plugin for Caliper to measure the latency and throughput results described in this paper. Caliper is now incubated and is part of the Hyperledger project \cite{hyperledger-caliper}.  Other works in this field are focussed on improving the security, performance and scaling issues of public blockchains such as Bitcoin and Ethereum  and other cryptocurrency platforms \cite{bitcoin-ng, algorand, powblockchains, securesharding, efficientconsensus}.  The design challenges in public platforms are quite different from the  involved in building permissioned platforms and therefore direct comparison of performance between the two is not accurate.

\section{Conclusions}
\label{sec:conclusion}

In this paper, we studied the performance characteristics of the Quorum blockchain platform. Quorum by default offers RAFT and IBFT for crash and Byzantine fault tolerance respectively. While it scales linearly for all workloads for the transaction send rates that we test for (upto 2100 tx/sec), the primary difference is in the transaction latencies. While reads have the lowest latencies, latencies of null and write workloads are largely dependent on the block time parameter (as expected). In case of the RAFT consensus algorithm, the throughput of the system does not change much by lengthening the block time, however latencies increase proportionately.  RAFT and IBFT are comparable in terms of throughput except that  RAFT performs slightly better at higher input transaction rates above 1650 tx/sec and IBFT slightly over performs RAFT for lower transaction rates, quite contrary to our expectation. 

Private contracts in Quorum result in lower throughput at higher load on the system due to extra overhead involved in secure communication and encryption/decryption operations employed between peers for confidentiality.  The throughput is lower compared to public contracts when the input transaction rate increases beyond 600 tx/sec. The maximum achievable load on the system was 900 tx/sec beyond which the system failed to reach consensus. 

Through our micro-benchmarking experiments, we showed that transaction latencies increase with increasing number of reads and writes in the smart contract as expected but showed that write latencies increase significantly when write set is as large as 1 million (Section~\ref{sec:rwsetsize}). Quorum provides a maximum payload size for transactions and events of 32KB for applications. Transaction latencies increase by 22\%-25\% when the transaction or event payloads are increased from 1 KB to 30 KB (Section~\ref{sec:payloadsizes}). Read and write latencies are relatively unaffected by the size of the data stored in the smart contract (Section~\ref{sec:keyvaluestoresize}). Therefore application developers can have considerable flexibility while choosing data set sizes and increasing key-value store sizes over time for applications are unlikely to degrade the performance of read and write operations. 



\begin{thebibliography}{1}

\bibitem{fabricpaper}
E.~{Androulaki}, A.~{Barger}, V.~{Bortnikov}, C.~{Cachin}, K.~{Christidis},
  A.~{De Caro}, D.~{Enyeart}, C.~{Ferris}, G.~{Laventman}, Y.~{Manevich},
  S.~{Muralidharan}, C.~{Murthy}, B.~{Nguyen}, M.~{Sethi}, G.~{Singh},
  K.~{Smith}, A.~{Sorniotti}, C.~{Stathakopoulou}, M.~{Vukoli{\'c}}, S.~{Weed
  Cocco}, and J.~{Yellick}.
\newblock {Hyperledger Fabric: A Distributed Operating System for Permissioned
  Blockchains}.
\newblock {\em ArXiv e-prints}, January 2018.
\newblock \url{https://arxiv.org/abs/1801.10228}


\bibitem{ethereum}
Vitalik Buterin.
\newblock Ethereum: A next-generation smart contract and decentralized
  application platform.
\newblock \url{https://github.com/ethereum/wiki/wiki/White-Paper}, 2014.
\newblock Accessed: 2016-08-22.

\bibitem{raft-quorum}
Raft-based consensus for Ethereum/Quorum
\newblock \url{https://github.com/jpmorganchase/quorum/blob/master/raft/doc.md}
\newblock [Online; accessed Apr 15, 2018].

\bibitem{raft}
In Search of an Understandable Consensus Algorithm - Diego Ongaro and John Ousterhout, Stanford University, 2014
\newblock \url{https://raft.github.io/raft.pdf}
\newblock [Online; accessed Apr 15, 2018].

\bibitem{paxos}
Paxos made simple - Leslie Lamport
\newblock \url{https://www.microsoft.com/en-us/research/uploads/prod/2016/12/paxos-simple-Copy.pdf}
\newblock [Online; accessed Apr 15, 2018].

\bibitem{raft-failure}
Raft-based consensus failure issue 
\newblock \url{https://github.com/coreos/etcd/issues/8553}
\newblock [Online; accessed Apr 15, 2018].

\bibitem{ibft}
Istanbul BFT
\newblock \url{https://github.com/ethereum/EIPs/issues/650}
\newblock [Online; accessed Apr 15, 2018].

\bibitem{privatetxflow}
Private Transaction Processing in Quorum
\newblock \url{https://github.com/jpmorganchase/quorum/wiki/Transaction-Processing#private-transaction-process-flow}
\newblock [Online; accessed Apr 15, 2018].

\bibitem{sawtoothlake}
Hyperledger Sawtooth Enterprise grade Platform
\newblock \url{https://sawtooth.hyperledger.org/docs/}
\newblock [Online; accessed Apr 15, 2018].

\bibitem{hyperledger-caliper}
Hyperledger Caliper
\newblock \url{https://www.hyperledger.org/projects/caliper}
\newblock [Online; accessed Apr 15, 2018].

\bibitem{quorum}
J.~P.~Morgan Chase.
\newblock {A Permissioned Implementation of Ethereum}.
\newblock \url{https://github.com/jpmorganchase/quorum}, 2018.
\newblock [Online; accessed Feb 20, 2018].

\bibitem{blockbench}
Tien Tuan~Anh Dinh, Ji~Wang, Gang Chen, Rui Liu, Beng~Chin Ooi, and Kian{-}Lee
  Tan.
\newblock {BLOCKBENCH:} {A} framework for analyzing private blockchains.
\newblock In {\em Proceedings of the 2017 {ACM} International Conference on
  Management of Data, {SIGMOD} Conference 2017, Chicago, IL, USA, May 14-19,
  2017}, pages 1085--1100, 2017.


\bibitem{bitcoin-ng}
Ittay Eyal, Adem~Efe Gencer, Emin~G\"{u}n Sirer, and Robbert Van~Renesse.
\newblock Bitcoin-ng: A scalable blockchain protocol.
\newblock In {\em Proceedings of the 13th Usenix Conference on Networked
  Systems Design and Implementation}, pages 45--59, Berkeley, CA, USA, 2016.
  USENIX Association.

\bibitem{fabric}
Hyperledger Fabric.
\newblock {Hyperledger Fabric}.
\newblock \url{https://www.hyperledger.org/projects/fabric}, 2017.
\newblock [Online; accessed Nov 14, 2017].

\bibitem{hyperledger}
Linux Foundation.
\newblock Linux foundation hyperledger project.
\newblock \url{https://www.hyperledger.org/}, 2017.
\newblock [Online; accessed Nov 14, 2017].

\bibitem{powblockchains}
Arthur Gervais, Ghassan~O. Karame, Karl W\"{u}st, Vasileios Glykantzis, Hubert
  Ritzdorf, and Srdjan Capkun.
\newblock On the security and performance of proof of work blockchains.
\newblock In {\em Proceedings of the 2016 ACM SIGSAC Conference on Computer and
  Communications Security}, CCS '16, pages 3--16, New York, NY, USA, 2016. ACM.

\bibitem{algorand}
Yossi Gilad, Rotem Hemo, Silvio Micali, Georgios Vlachos, and Nickolai
  Zeldovich.
\newblock Algorand: Scaling byzantine agreements for cryptocurrencies.
\newblock In {\em Proceedings of the 26th Symposium on Operating Systems
  Principles}, SOSP '17, pages 51--68, New York, NY, USA, 2017. ACM.

\bibitem{multichain}
Dr.~Gideon Greenspan.
\newblock Multichain private blockchain - white paper.
\newblock \url{https://www.multichain.com/download/MultiChain-White-Paper.pdf},
  2016.
\newblock Accessed: 2018-02-20.

\bibitem{corda}
Mike Hearn.
\newblock Corda: A distributed ledger.
\newblock \url{https://docs.corda.net/_static/corda-technical-whitepaper.pdf},
  2016.
\newblock Accessed: 2018-02-20.

\bibitem{securesharding}
Loi Luu, Viswesh Narayanan, Chaodong Zheng, Kunal Baweja, Seth Gilbert, and
  Prateek Saxena.
\newblock A secure sharding protocol for open blockchains.
\newblock In {\em Proceedings of the 2016 ACM SIGSAC Conference on Computer and
  Communications Security}, CCS '16, pages 17--30, New York, NY, USA, 2016.
  ACM.


\bibitem{chain}
Chain Enterprise grade Blockchain Platform
\newblock \url{https://chain.com/}, 2018.
\newblock [Online; accessed Apr 15, 2018].

\bibitem{parity}
Parity - Ethereum Client
\newblock \url{https://www.parity.io/}
\newblock [Online; accessed Apr 15, 2018].


\bibitem{bitcoin}
Satoshi Nakamoto.
\newblock Bitcoin: A peer-to-peer electronic cash system.
\newblock \url{https://bitcoin.org/bitcoin.pdf}, Dec 2008.
\newblock Accessed: 2015-07-01.

\bibitem{efficientconsensus}
Rafael Pass and Elaine Shi.
\newblock {Hybrid Consensus: Efficient Consensus in the Permissionless Model}.
\newblock In Andr{\'e}a~W. Richa, editor, {\em 31st International Symposium on
  Distributed Computing (DISC 2017)}, volume~91 of {\em Leibniz International
  Proceedings in Informatics (LIPIcs)}, pages 39:1--39:16, Dagstuhl, Germany,
  2017. Schloss Dagstuhl--Leibniz-Zentrum fuer Informatik.


\bibitem{caliper}
Huawei Technologies.
\newblock {Caliper: A Blockchain Benchmark framework}.
\newblock \url{https://github.com/Huawei-OSG/caliper/}, 2017.
\newblock [Online; accessed Nov 14, 2017].

\bibitem{huawei}
Huawei Technologies.
\newblock {Huawei Technologies}.
\newblock \url{http://www.huawei.com/en/}, 2017.

\bibitem{constellation}
Constellation.
\newblock {Constellation - A self-managing peer-to-peer system}.
\newblock \url{https://github.com/jpmorganchase/constellation}.
\newblock [Online; accessed Mar 12, 2018].

\bibitem{pbft}
Miguel Castro and Barbara Liskov.
\newblock Practical byzantine fault tolerance.
\newblock In {\em Proceedings of the Third Symposium on Operating Systems
  Design and Implementation}, OSDI '99, pages 173--186, Berkeley, CA, USA,
  1999. USENIX Association.


\end{thebibliography}

\end{document}